\documentclass[journal]{IEEEtran}

\usepackage{cite}
\usepackage[utf8]{inputenc} 
\usepackage[english]{babel}
\usepackage{todonotes}
\usepackage{color,soul} 

\usepackage[square,numbers]{natbib}
\usepackage[hyphens]{url}
\usepackage{booktabs}
\usepackage{enumitem}
\usepackage{pdflscape}  
\usepackage{afterpage} 
\usepackage{float}  
\usepackage{placeins} 
\usepackage{graphicx}
\usepackage{pgfgantt}
\usepackage{xcolor}
\usepackage[normalem]{ulem} 

\usepackage{amssymb} 
\usepackage{multirow}
\usepackage{multicol}
\usepackage{colortbl}
\usepackage{numprint}
\usepackage{textcomp}
\usepackage{amsmath}
\usepackage{tikz}
\usepackage{array}
\usepackage{wrapfig}
\usepackage{algorithm}
\usepackage{algorithmic}
\usepackage{caption}
\usepackage{placeins}
\usepackage{balance}
\begin{document}
\title{Wireless Connectivity of a Ground-and-Air Sensor Network}

\author{Clara R. P. Baldansa,
        Roberto C. G. Porto, 
        Bruno J. Olivieri de Souza, 
        Vítor G. Andrezo Carneiro,
        Markus Endler
        }

\maketitle

\begin{abstract}

This paper shows that, when considering outdoor scenarios and wireless communications using the IEEE 802.11 protocol with dipole antennas, the ground reflection is a significant propagation mechanism. This way, the Two-Ray model for this environment allows predicting, with some accuracy, the received signal power. This study is relevant for the application in the communication between overflying Unmanned Aerial Vehicles (UAVs) and ground sensors. In the proposed Wireless Sensor Network (WSN) scenario, the UAVs must receive information from the environment, which is collected by sensors positioned on the ground, and need to maintain connectivity between them and the base station, in order to maintain the quality of service, while moving through the environment.

\end{abstract}

\begin{IEEEkeywords}
IoT, WSN, UAV, Mesh, Two-Ray Model, Wireless Channel Prediction
\end{IEEEkeywords}

%
\IEEEpeerreviewmaketitle

\section{Introduction}

\IEEEPARstart{D}{ata} collection from wireless sensor networks (WSN) on the ground by unmanned aerial vehicles (UAVs) has been widely researched in the last decades and has led to some specific solutions for environmental monitoring, security, precision agriculture, and several other applications \cite{Goudarzi2019,Popescu2020}. Common to all these applications is that the geographic region to be monitored or controlled is either difficult to access, is very large, or is hazardous, making overflying the only feasible way to collect the data.

In parallel, the IoT industry has propelled the miniaturization of System-on-Chips, short-range low-power wireless communications, mesh network technologies, and battery-operated sensors and microcontrollers, making it possible to remotely sense almost any environment or physical space \cite{Ali2022}. However, as long as node mobility is a key enabler of remote sensing, such as with UAVs, there are still huge challenges to be solved, such as intermittent wireless connectivity, radio interference, choice of antennas, fading, and handoff. 

In previous research, our group proposed a distributed algorithm for UAV flight coordination and cooperative sensor data collection, called Distributed Aerial Data Collection Algorithm (DADCA) \cite{Olivieri-JISA2020}. It permits an arbitrary and dynamic set of UAVs to collaborate with each other, self-organizing themselves to collect and haul data from desired points of interest (POI) to a base station. In parallel, focusing on effective wireless communication in a ground-based mesh network, we proposed the Mobility Aware Mesh routing protocol (MAM) \cite{Paulon-Winsys2021}, which is an alternative protocol to the Bluetooth Mesh standard (BT Mesh), focusing on mobile sink/collectors.

In this context, the Ground-and-Air Dynamic Sensor network (GrADyS) project was conceived. This project aims, among other objectives, to test the interaction and interoperability between dynamic mesh network protocols, in order to explore the benefits of a full air-with-ground mesh network. One important point of the GrADyS project is the validation of the air-with-ground interactions and protocols, not only through network simulators but also in real-world field experiments, i.e., considering realistic problems, such as the wireless propagation environment.

\begin{figure}[ht!]
    \includegraphics[width=4.5cm]{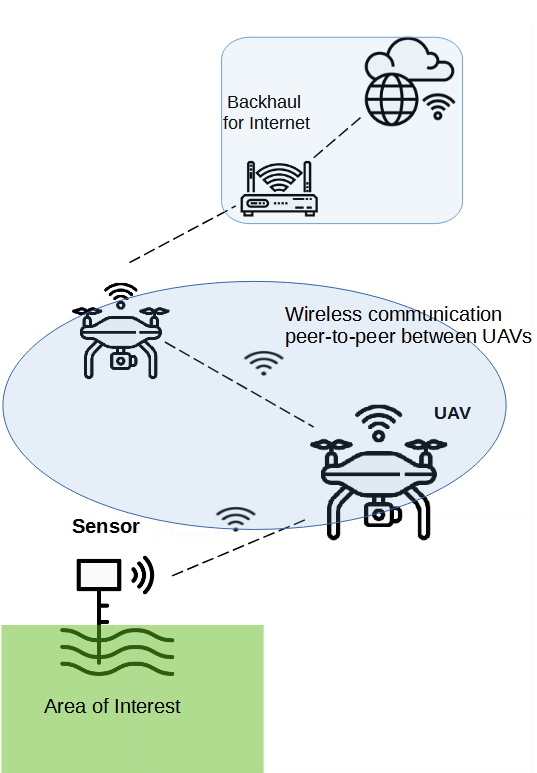}
    \centering
    \caption{GrADyS project scheme.}
    \label{fig:GrADyS}
\end{figure}

Figure \ref{fig:GrADyS} shows a network of UAVs, that collect information from sensors, located in the area of interest, through peer-to-peer wireless communication. The UAVs also exchange information with each other and with the central station, which works as a backhaul to the Internet.

\vspace{-0.2cm}

Many works disregard antenna characteristics and consider the transmission range as a circle, i.e., the transmitted power of a node decays with the square of the distance \cite{Goudarzi2019,Popescu2020,Paulon2022}. Thus, this paper aims to predict the wireless communication behavior between the sensors and the UAVs, more specifically, to predict the received signal power, based on transmitter power, type and polarization of antennas, electromagnetic reflectivity of soil, and characteristics of a realistic radio transmitter.

This paper is structured as follows. Section 2 presents the motivation of the project and the problem to be solved: model the wireless propagation channel at the Wi-Fi frequency, in order to simulate the communication between terrestrial sensors and UAVs that fly over the area to be inspected. In Section 3, we discuss the Two-Ray Model, a specific case of the Ray Tracing algorithm. Sections 4 and 5 are intended to present the methodology used in the experiments, as well as the antenna measurements. Afterward, the results obtained experimentally are shown and a comparison is made with the proposed model. Finally, Section 6 makes the conclusions.



\section{Wireless Connectivity for Outdoor Scenarios}
\label{sec:problem}

As mentioned before, energy-efficient techniques and greater processing power are critical aspects of IoT industry applications [3]. Without these technological advances, WSN would not be possible. These networks use devices that sense, collect and transmit data to overflying UAVs, which send the collected information to a central station that serves as a backhaul. As a result, several works seek to determine the best pre-calculated path, followed by drones, that minimizes energy consumption and ensure that data packets have been collected successfully  \cite{Goudarzi2019, Popescu2020, TSP}.

However, these works attempt to determine the most optimized trajectory using computational models, that consider the transmission range between UAVs and the ground sensors within a circular region. In other words, they use in their modeling an isotropic ideal radiator and a decay of the electromagnetic energy with the square of the distance. Our work has a slightly different approach, aiming to optimize the UAVs' path based on efficient data collection. That is, we seek to estimate the points of the UAVs trajectory in which the Received Signal Strength (RSS) is greater, in the sense of where there is a better Quality of Service (QoS) of the signal, considering the influence of the antenna radiation pattern and the influence of reflections in the wireless channel.

The characteristics of the antennas and their applications in wireless communication \cite{WU2022374} and the study of the wireless channel itself \cite{https://doi.org/10.48550/arxiv.2202.09740} are targets of special interest by the research community. The objective is to improve the wireless communication quality and to be able to predict the wireless channel signal strength at different points of the terrain, through a database of some measurements and the use of mathematical and computational models, without the need for costly work of measuring the power values at each of these points in loco.

In this context of channel modeling, one of the most known propagation loss mechanisms, that limits the communication range, is the Free-Space Path Loss (FSPL). This model does not consider any reflection effect or the presence of obstacles and its loss is given by:

\begin{equation}
FSL(dB) = 32.45 + 20\log(l) + 20\log(f),
\end{equation}

\noindent
where $l$ is the Line-Of-Sight (LOS) distance, in meters, and $f$ is the signal frequency, in gigahertz.

In this work, we compare the experimental data of the power received by a sensor, as the UAV moves through the air, with the FSPL model, where the power also decays with the square of the distance, but the frequency of the signal is considered. However, it is known that only a LOS ray between the transmitter and the receiver is hardly the only propagation mechanism. Hence, the FSPL model alone becomes, in most situations, inaccurate for describing the signal propagation.

Ray Tracing is a channel modeling method to determine the rays that leave a transmitter and arrive at the receiver, considering a finite number of reflectors and their dielectric properties. When considering outdoor scenarios, the Ray Tracing technique becomes the Two-Ray model, a particularization in which two rays are considered: the LOS and the one that comes from ground reflection. Thus, in this scenario, there are no obstacles that cause other reflections or diffraction \cite{Molisch2004}.

In related work, some authors suggest that, for long-distance Wi-Fi signals, it is possible to consider only one ray reaching the receiver, and thus, model the channel using only FSPL. However, they admit restrictions to this method and, for certain scenarios, suggest the use of other deterministic models \cite{paper2}. Other authors tried to overcome the limitations of FSPL, by formulating their own proposed path loss equation, which takes into account the heights of the antennas in the calculations. They concluded that their path loss equation is more accurate than those normally used \cite{paper3}.

As pointed out by \cite{paper4}, the biggest limitation when using the FSPL model is the height of the antennas. If the ratio between the heights of the antennas and the distance between them is small, then the influences of the Earth's curvature and the Fresnel zone cannot be neglected. This is our case since the sensors should be dropped by UAVs and land very close to the ground. Another article also considered the sea surface reflections important, in the case of an air-to-ground channel over the sea surface, especially when the antenna altitude is low \cite{paper5}. In this context, according to \cite{paper4}, the estimates of RSS using the Two-Ray model are far more precise than those made using the FSPL equation.
\section{Two Ray Model}

Wireless propagation channels have the disadvantage, in relation to guided channels, that the signal is subject to multipath. Hence, the electromagnetic waves can reach the receiver through different paths, being able to be reflected, refracted, or diffracted along the way.

As mentioned in Sec. \ref{sec:problem}, the Two-Ray model is the simplest case of Ray Tracing, which considers only two paths. Thus, the energy that arrives at the receiving antenna comes from different directions and at different times (delay). In turn, when arriving at the receiver, the waves can add constructively or destructively, depending on their phase difference.

The mathematical formulation of the Two-Ray model is:
\begin{equation}
P_r=P_t\left[\frac{\lambda}{4\pi}\right]^2\left[\frac{\sqrt{G_l}}{l}+\frac{\rho_s D\Gamma\sqrt{G_r}e^{-j\Delta\phi}}{x+x'}\right]^2,
\end{equation}

\noindent
where $\lambda$ is the wavelength, $G_l$ is the total antenna gain of the LOS path, $\rho_s$ is the ground scattering coefficient, $D$ is the divergence factor of the Earth, $\Gamma$ is the Fresnel reflection coefficient of the ground, $G_r$ is the total antenna gain of the reflected path, $\Delta\phi$ is the phase difference between the two received rays, and $x+x'$ is the reflected path length. 

\begin{figure}[ht]
    \includegraphics[width=8cm]{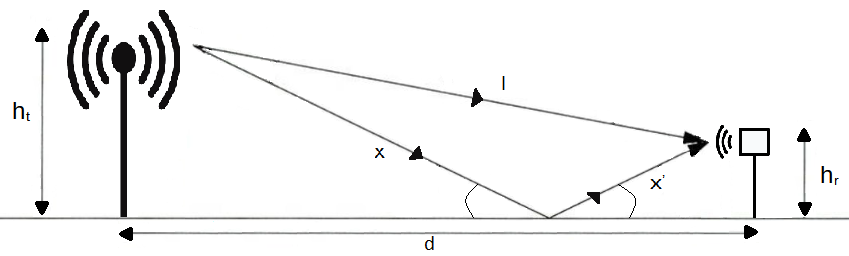}
    \centering
    \caption{Schematic of the Two-Ray model.}
    \label{fig:Schematic of 2 ray model}
\end{figure}

Reflection occurs when the propagating electromagnetic wave hits an obstacle with dimensions much larger than its wavelength. The intensity of the reflected wave is defined through the Fresnel reflection coefficients, which depend on the antenna polarization (vertical or horizontal), the angle of incidence of the wave with the ground, and the complex relative permittivity of the soil, which considers its conductivity. The soil parameters used in our simulations were estimated and are shown in Table \ref{tab:parameters}:




\begin{table}[ht]
\begin{center}
\caption{Parameters of the Wireless Channel Simulations}
\label{tab:parameters}
\vspace{0.5cm}
\begin{tabular}{c c c}

\toprule[1.0pt]
Parameter & Description & Value\\
\midrule[1.0pt]
$\varepsilon_r$ & relative permittivity grass & 42 \\
$\varepsilon_r$ & relative permittivity concrete & 1.7 \\
$\sigma$ & conductivity soil & 0 (mS/m) \\

\midrule[1.0pt]\\

\end{tabular}
\end{center}
\end{table}

Moreover, the flat Earth propagation model is admitted, where the divergence factor ($D$) is equal to 1. This consideration can be made since the antennas are small in relation to the radius of the earth and the distance between them is much greater than the order of size of their heights.

Furthermore, another effect to consider is the diffuse reflection of rays on rough ground. The scattering coefficient that quantifies the degree of scattering is given by:

\begin{equation}
    \rho_s=e^{-0.5\Delta\phi_1^2}, \Delta\phi_1 = \frac{4\pi\Delta h\cos{\theta_i}}{\lambda},
\end{equation}

\noindent
where $\Delta h$ is the standard deviation of the ground heights $\theta_i$ is the angle of incidence of the wave with the ground.

\section{Field Measurements}

This section discusses our methodology for analyzing path loss for links using the IEEE 802.11 protocol between two omnidirectional antennas, simulating wireless communication between the drone and the ground sensor. The software and hardware equipment, used to measure both the received and transmitted power, and the Two-Ray model experimental setup, used for the analysis, are described.

\begin{figure}[ht]
    \includegraphics[width=7cm]{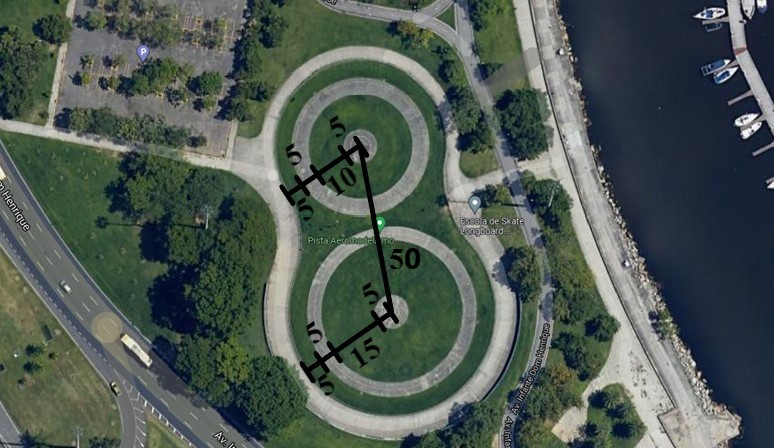}
    \centering
    \caption{Propagation test environment.}
    \label{fig:Aterro}
\end{figure}

Therefore, in order to simulate the propagation of RF waves between the drone and the sensor node on the ground, two ESP32 boards with a swivel whip antenna were employed, one attached to a 5-meter rod and the other at a height of 15 centimeters from the ground, respectively. In the experiment, the antenna that represents the sensor is kept fixed in one position to collect Wi-Fi signal magnitude measurements. The antenna that plays the role of a drone is moved away to different positions in the scenario shown in Figure \ref{fig:Aterro}.

\begin{figure}[ht]
    \includegraphics[width=6cm]{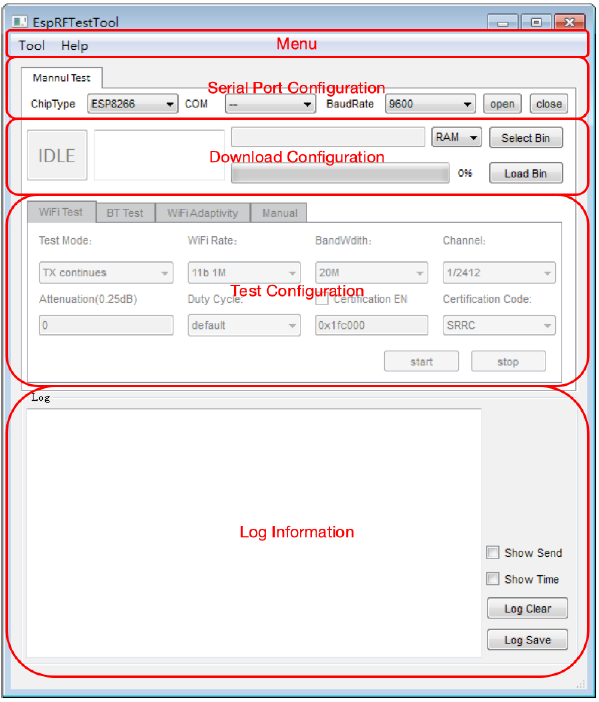}
    \centering
    \caption{EspRFTestTool main interface.}
    \label{fig:EspRFTestTool main interface}
\end{figure}

For the setup of the half-wave dipoles antenna parameters, it was employed the software ESPRFTestTool, developed by Espressif Systems. For the transmitting antenna,in the test configuration interface, the parameters necessary for the simulation were inserted in the WiFi test tab: the frequency of 2.412GHz (channel 1), the transmission power of 10 dBm and TX Tone mode which, according to the manufacturer’s specifications, is a “single carrier TX signals”, also known as carrier wave (CW). In the receiver, which simulates the sensors on the ground, the RX packet mode was set and the minimum receive power is -85 dBm. In figure \ref{fig:EspRFTestTool main interface} shows the graphical interface of the software ESPRFTestTool to program the ESP32 and adjust the antenna parameters.

In addition to the antenna settings, such as maximum gain, RF frequency and receiver sensitivity, the simulation considered the influence of other factors, such as losses in the cables and devices used. Table \ref{tab:parameters} shows the parameters used in the wireless channel simulations \cite{Ometov2019}. Another relevant aspect of the work using the half-wave dipoles antenna, which present omnidirectional planes of propagation, was the use of both polarization planes: vertical and horizontal.

\begin{table}[ht!]
\begin{center}
\caption{Parameters of the Wireless Channel Simulations}
\label{tab:parameters}
\vspace{0.5cm}
\begin{tabular}{c c c}

\toprule[1.0pt]
Parameter & Description & Value\\
\midrule[1.0pt]
$P_{TX}$ & Transmitter power & 10 dBm \\
$L_{T}$ & ESP32 transmitter loss & 1.5 dB \\
$L_{R}$ & Adalm Pluto receiver loss & 2.5 dB \\
$L_{C}$ & Cable loss & 0.6 dB \\
$L_{D}$ & Resistive divider loss & 7.1 dB \\
$G_{max}$ & Maximum antenna gain & 1.97 dB \\
$S_{r}$ & ESP32 receiver sensitivity  & --85 dBm \\
$f$ & Signal frequecy & 2.412 GHz\\
$d$ & Link horizontal distance & 0 -- 50 m \\
\midrule[1.0pt]\\

\end{tabular}
\end{center}
\end{table}

\begin{figure}[ht]
    \includegraphics[width=8cm]{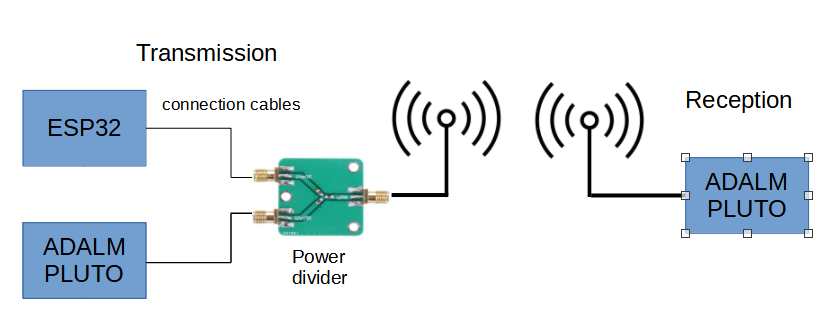}
    \centering
    \caption{Setup of the transmitter and receiver.}
    \label{fig:Setup Transmitter}
\end{figure}

\begin{figure}[ht]
    \includegraphics[width=8cm]{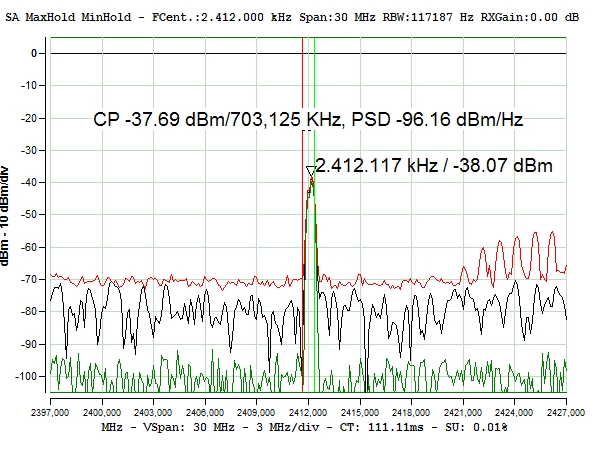}
    \centering
    \caption{Measurement made with Satsagen.}
    \label{fig:Satsagen}
\end{figure}

To read the power measurements is used the software defined radio (SDR) Adalm Pluto combined with the windows application Satsagen, figure \ref{fig:Satsagen}, that allows the SDR to be used as a spectrum analyzer. In the transmitter, to be sure of the power that was effectively sent to the receiver, the reading is also made using the spectrum analyzer, the configuration framework can be seen in figure \ref{fig:Setup Transmitter}, where the signal strength divider is used to split up the transmitted signal of the ESP32 between the antenna and the Adalm Pluto.

Another important parameter in the simulation of the Two-Ray model is the type and radiation characteristics of the antenna used, directly influencing the mathematical expression through the antenna gain for a given angle. In the next section, we will show the direction of the power propagation through the plot the radiation pattern.
\section{Antenna Measurements}
\label{sec:antenna}

Since antennas make the connection between transmitter and receiver through the wireless propagation channel, they have major importance in the analysis of the wireless link. Thus, its characteristic parameters, such as radiation pattern, gain, bandwidth, polarization, and impedance, have a great impact on a wireless connection. By the principle of reciprocity, it is considered that the characteristics of antennas are the same regardless of whether they are used for signal transmission or reception.
	
An antenna's directivity defines its ability to focus radiated energy in a given direction.
The gain of an antenna is related to its directivity, but it also takes into account the antenna losses, quantified through its efficiency. The antenna radiation pattern is the graphical representation of the irradiation intensity as a function of the propagation direction angle, defined for the far field.

\begin{figure}[ht]
    \includegraphics[width=8cm]{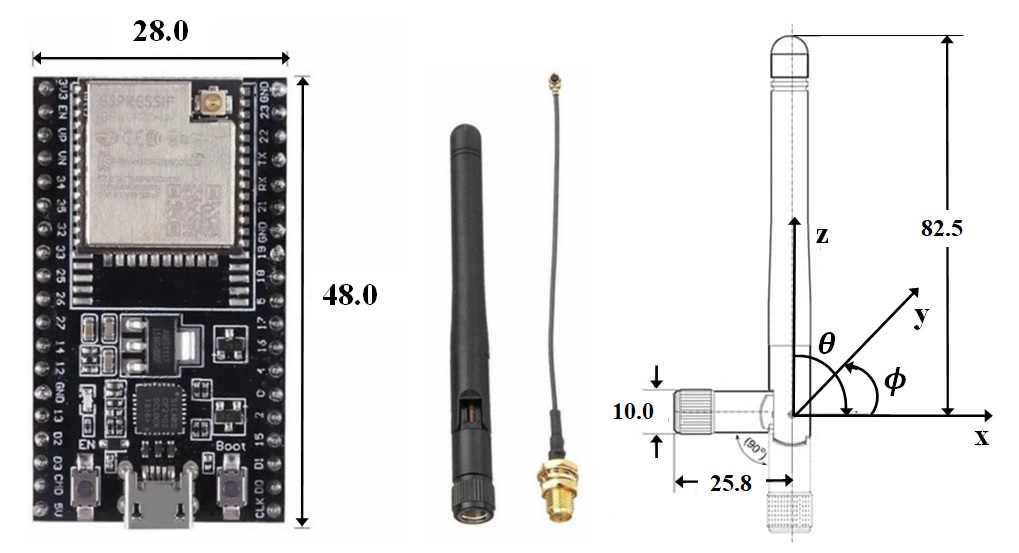}
    \centering
    \caption{ESP32-WROOM-32U and its dipole antenna with dimensions (in mm).}
    \label{fig:ESP-Dipole}
\end{figure}

\begin{figure}[ht]
    \includegraphics[width=9cm]{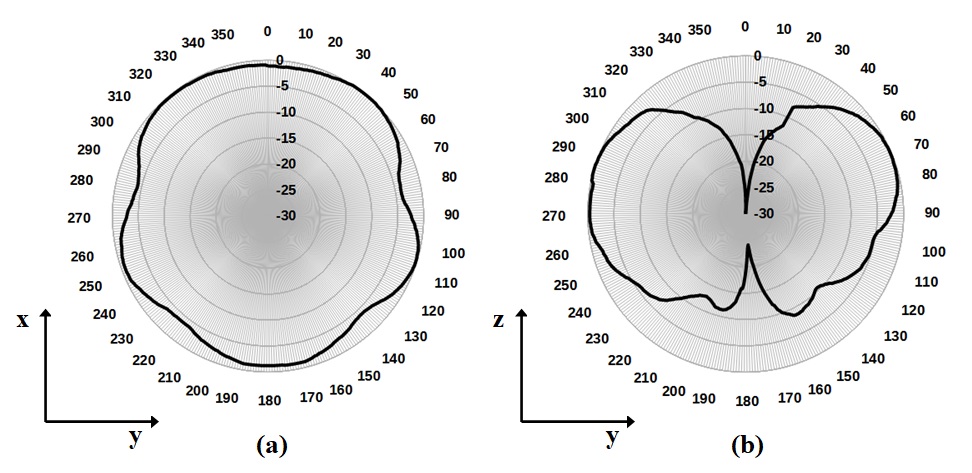}
    \centering
    \caption{Dipole antenna diagram: (a) xy plane ($\phi$ in degrees) and (b) yz plane ($\theta$ in degres).}
    \label{fig:DipoleDiagram}
\end{figure}

For the calculation of the project, the far-field pattern of an external half-wave dipole antenna, attached to an ESP32 module, as shown in Figure \ref{fig:ESP-Dipole}, was measured in an anechoic chamber, made with absorbing material. These measurements are shown in Figure \ref{fig:DipoleDiagram}.

\section{Results and Discussions}

This section provides the results of field experiments performed in an outdoor area without obstacles in the proximity of the antenna in order to compare the measurements obtained with the theoretical curves traced according to the Two-Ray model and the FSPL model in our Matlab program. We want to know if the Two-Ray model approach is valid to characterize the wireless channel for the IEEE 802.11 standard for outdoor links. Another objective is to analyze which antenna polarization (vertical or horizontal) is the most appropriate for the link between sensor nodes and overflying UAVs.

\begin{figure}[ht!]
    \includegraphics[width=10cm]{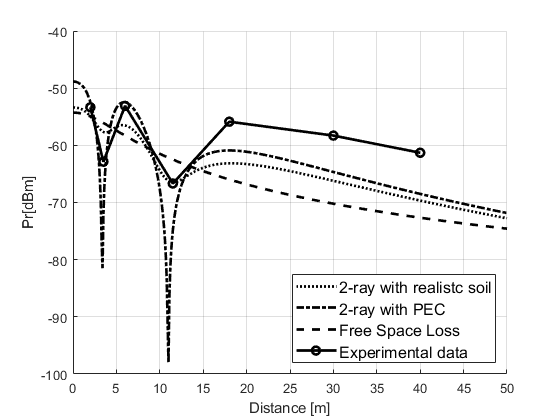}
    \centering
    \caption{Received signal power for horizontal polarization.}
    \label{fig:polH}
\end{figure}

\begin{figure}[ht!]
    \includegraphics[width=9cm]{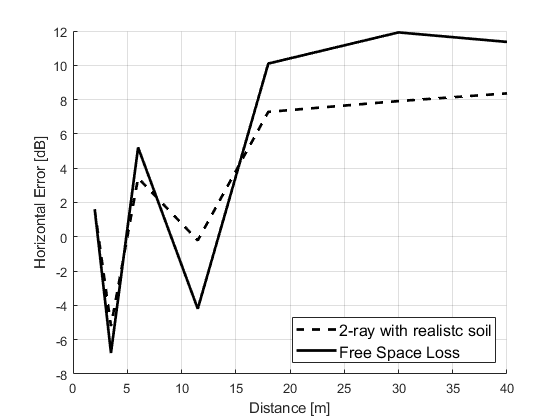}
    \centering
    \caption{Error for horizontal polarization.}
    \label{fig:polHerro}
\end{figure}

The dash-dotted curve in Figure \ref{fig:polH} refers to the simplified theoretical curve of the two-ray model where the following considerations are made: the ray is fully reflected with phase inversion, that is, $\Gamma = -1$, specular reflection is considered, that is, $\rho_s = 1$ and therefore there is no loss of energy by scattering the rays and the effect of the Earth ray $D = 1$ is neglected. Although this modeling does not represent all the propagation mechanisms present in a real environment, this curve shows the theoretical maximum and minimum values of power at the receiver. While the dotted curve represents the modeling using the equations presented and considering the real ESP32 antenna, which has slightly different radiation characteristics from a dipole antenna described in the literature, as seen in the section \ref{sec:antenna} and also and considers a realistic soil with diffuse reflection losses in the soil. The dashed curve indicates the model well-known Free Space Path Loss, and the the curve with circular markers shows the experimental measurements.

\begin{figure}[ht!]
    \includegraphics[width=10cm]{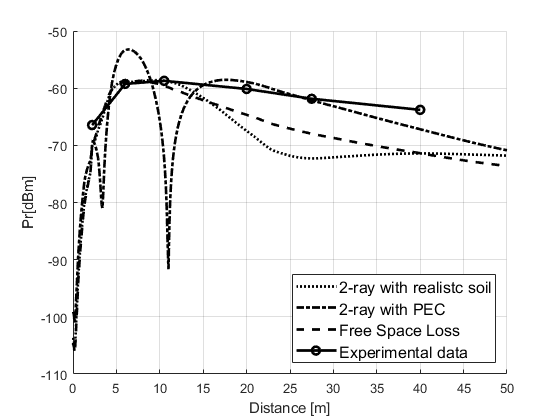}
    \centering
    \caption{Received signal power for vertical polarization.}
    \label{fig:polV}
\end{figure}

\begin{figure}[ht!]
    \includegraphics[width=9cm]{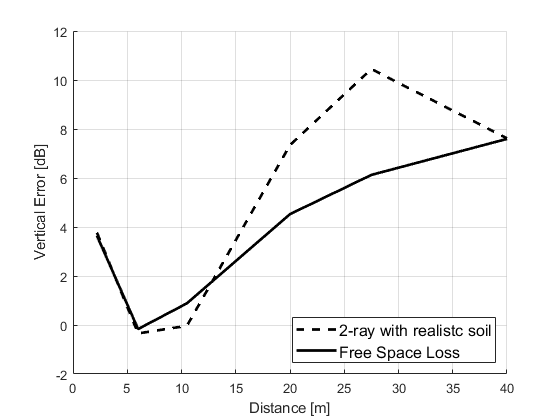}
    \centering
    \caption{Error for vertical polarization.}
    \label{fig:polVerro}
\end{figure}

Figure \ref{fig:polHerro} shows the error curves in dB calculated from the difference of the values measured in the field and the curve of the two-ray model using the ESP32 (dashed line) and the difference of the measured values for the curve of the FSPL model (solid line). The same reasoning and color legend are used for the antenna with vertical polarization, as can be seen in the figures
\ref{fig:polV} and \ref{fig:polVerro}.

\section{Conclusion}

Analyzing the results obtained through experiments in outdoor for a wireless link using the IEEE 802.11 standard on channel 1 and comparing with the simulation of the two-ray model and FSPL in the Matlab software, both for the horizontal and vertical polarization of the antenna dipole, we conclude that, for horizontal polarization, the two-ray model was more accurate in determining the Received Signal Strength.

 Therefore, we can say that the phenomenon of reflection on the ground, which leads to constructive or destructive interference when reaching the receiver, is a suitable modeling for the context of communication between UAVs and the sensor on the ground in an open environment.
 
We emphasize the better accuracy of the modeling using the horizontal polarization in relation to the vertical, because comparing the \ref{fig:polHerro} and \ref{fig:polVerro} figures, it is noted that in the first figure the error related to the curve of the ESP32 two-ray model is always below the error of the curve by FSPL modeling. But, for our test environment we know that there will be a reflection on the ground, consequently the two-ray model should present better results in relation to the prediction of the received power, so we can conclude that there are other rays that were not considered and that in the diffraction zone these rays add constructively making the RSS found to be greater than predicted by our theoretical two ray model.

In conclusion, it is observed that for distances smaller than 20 meters (interference zone) the simulation of the two-ray model fits well to model the wireless propagation channel according to the experimental data, but for greater distances between the two antennas, approximately for distances greater than 20 meters (diffraction zone), the received power intensity is around 10dBm above the theoretical curve expected according to our theoretical model. Thus, for future work, we want to consider the influence of other scattered rays from non-specular reflection, which can reach the receiver in random directions with different energies and that add up constructively in this region. This approach has already been studied in other works on propagation of electromagnetic waves applied to mobile communications \cite{diffuse} and indicated a conclusion that the modeling using optical ray launch technique can have its accuracy improved if we consider the significant contribution of diffuse scattering rays.


\section*{Acknowledgment}

This study was financed in part by AFOSR grant FA9550-20-1-0285.

\ifCLASSOPTIONcaptionsoff
  \newpage
\fi


\bibliographystyle{unsrt}
\bibliography{bib-olivieri.bib, bib-markus.bib, bib-andrezo.bib,  bib-matuchewski.bib, bib-clara.bib,bib-Machado.bib,bib-relatedwork.bib, bib-conclusion.bib}







\end{document}